\documentstyle[prcrc,epsfig,11pt,fleqn]{article}

\newcommand{\be}{\begin{equation}} 
\newcommand{\ee}{\end{equation}}

\newcommand{\bea}{\begin{eqnarray}} 
\newcommand{\eea}{\end{eqnarray}} 
\newcommand{\eg}{{\it e.g., }} 
\newcommand{\ie}{{\it i.e., }} 
\newcommand{\OM}{\Omega_m}

\newcommand{\OL}{\Omega_{\Lambda}}

\newcommand{\mz}{$m$-$z$\ }

\title{Inhomogeneities and the Intergalactic Distance-Redshift Relation} 
\author
{R.  Kantowski 
\address{Department of Physics \& Astronomy\\
University of Oklahoma\\ 
Norman, OK 73019, USA }
}

\begin{document}
\maketitle

\begin{abstract}  
Modifications in Friedmann-Lema\^itre-Robertson-Walker
(FLRW) Hubble diagrams caused by  mass density inhomogeneities 
are used to illustrate possible effects on a
determination of the mass parameter $\OM$ and the cosmological constant $\Lambda$.
The values of
these parameters inferred from a given set of observations depend on the fractional amount of
matter in inhomogeneities and can  differ from those obtained by using
standard FLRW Hubble diagrams by as much as a factor of two.

\end{abstract}

\section{Introduction} 
The pressure-free FLRW models of GR 
have, for a very long time,  been assumed to adequately describe the large scale geometry of 
our universe but only recently have observational techniques  become available which 
promise to determine values for all three of their requisite parameters $\{H_0,\ \OM$, and $\OL\equiv
{\Lambda c^2/(3H_0^2)}\}$ and 
implicitly test this assumption.  As an example,  corrected magnitudes and redshifts ($m$-$z$) 
for Type Ia supernovae (SNe Ia) are
measured, plotted, and compared with theoretical $m(H_0,\ \OM,\OL; z)$ curves computed for the FLRW
models \cite{PS1,PS2,GP}. Because these models are isotropic and homogeneous, and
our universe appears quite inhomogeneous, modifications in the FLRW predictions have long been proposed
and estimated \cite{YZ,KR}. When a wide angle measurement  of the CMB is made the average mass density
of the universe (the FLRW value for $\rho_0$) likely exists within the radio beam collected by the antenna.
However, when small objects such as 
SNe Ia are observed at $z< 1$, a mass density significantly less than 
the average is `likely' to be in the observing optical path. 
In particular, if the underlying mass density approximately follows luminous matter
(\ie associated with bounded galaxies) then effects of a diminished mass density in the observing beams 
on relations like
$m(\OM,\OL;z)$ are important.  The majority of currently observed SNe Ia are not
being seen through foreground galaxies and whether or not this is due to selection
(rather than statistics) is not important.  If the objects observed do not have the average FLRW
mass density $\rho_0$ in their foregrounds then the FLRW \mz\ relation does not apply to them.
Ultimately some SNe Ia should exist behind foreground galaxies and for these, \mz\ should be
computed using the lensing formulas.  These formulas \cite{BR,CJ} contain 
source-observer,
deflector-observer, and source-deflector distances, respectively $D_s, D_d$, and $D_{ds}$, all of
which depend on the mass density in the observing beam, {\bf excluding} the deflector.
These distances will not be given by the standard FLRW result if the observing beam contains less
than the average FLRW mass density, but instead given by the `intergalactic' distance discussed here.

\section{Inhomogeneous Optics}

Hubble diagrams (or equivalently distance-redshift relations) for inhomogeneous universes are found by 
integrating the appropriate equation for the cross sectional area $A(z)$ of a beam propagating through them 
(see \cite{KR,KVB,KR2}).
When the only inhomogeneities are galaxies and their converging/lensing effects are taken into account by lensing 
formulas as 
indicated above, the appropriate  
equation to solve is:
\bea 
&&(1+z)^3\sqrt{1+\OM z+\OL[(1+z)^{-2}-1]}\times\nonumber\\ 
&&\hskip 1 in {d\ \over
dz}(1+z)^3\sqrt{1+\OM z+ \OL[(1+z)^{-2}-1]}\,{d\ \over dz}\sqrt{A(z)}\nonumber\\ 
&&\hskip 2.0 in
+ {(3+\nu)(2-\nu)\over 4}\OM(1+z)^5\sqrt{A(z)}=0,  \label{Ab0} 
\eea
where the parameter $\nu$ is defined by the fraction of matter ${\rho_I / \rho_0}={\nu(\nu+1)/ 6}$ 
in the universe existing in inhomogeneities 
and excluded from the observing beams (\eg galactic matter). The luminosity distance-redshift relation is given by
integrating (\ref{Ab0}) from source to observer and evaluating $D_{\ell}^2\equiv 
{(A\big|_0 / \delta\Omega)}(1+z)^2$, \ie the area at the observer divided by the solid angle at the source
and multiplied by two factors of the redshift. Eqn. (\ref{Ab0}) was recently recognized \cite{KR2} as a Lame$^\prime$ equation
and analytic solutions were given in terms of Heun functions. Because this $D_{\ell}(z)$
comes from neglecting optical effects of galactic mass it
might be reasonably dubbed the `intergalactic' luminosity distance-redshift relation thus giving on a log scale, 
the `intergalactic' Hubble diagram. 
\section{Errors Made by Using the Wrong Hubble Diagram}
With a little effort the reader can see from (\ref{Ab0}) how increasing from the FLRW value of $\nu= 0$ 
(100\% homogeneous)  to $\nu= 2$ (100\% clumpy)
increases the area at observation and hence decreases the apparent luminosity of an object. If the wrong
Hubble diagram is used, incorrect values of two of the three cosmic parameters will be obtained. Figures 
1 and 2 are intended to illustrate just how large an error can be made. 

If the distance modulus of a source such as SN 1997ap at $z=0.83$  were {\bf precisely} known 
(\eg see the two sample horizontal lines) then  a determination of  $\OM$ 
could be made from Figure 1, 
assuming $\OL$ were somehow known; or from Figure 2, a determination of $\OL$ could be made 
if $\OM$ were somehow known.  From Fig. 1 the reader can easily see that
the determined value of $\OM$ depends on the clumping parameter $\nu$. The $\OM$ value 
will be about 95\% larger for a $\nu=2$ completely clumpy universe than it will be 
for a $\nu=0$ completely smooth FLRW  universe. Equivalently, $\OM$ could be 
underestimated by as much as 50\% if FLRW is used. The maximum underestimate is reduced to  33\% 
 at the smaller redshift of $z=0.5$ (see a similar result for $\OL=0$ in \cite{KVB}).
These conclusions are not sensitive to the value of $\OL$.

Slightly different conclusions follow from  Fig. 2 about $\OL$. 
The discrepancy in the determined value of $\OL$ is $\Delta \OL \sim -0.14$   
for $\nu=2$ compared to $\nu=0$, and is not sensitive to the 
distance modulus.  The  discrepancy is halved, $\Delta \OL \sim -0.07$, at 
a smaller redshift of $z=0.5$. 
\begin{figure}[htb]
\centerline{
\begin{minipage}[t]{120mm}
\epsfig{figure=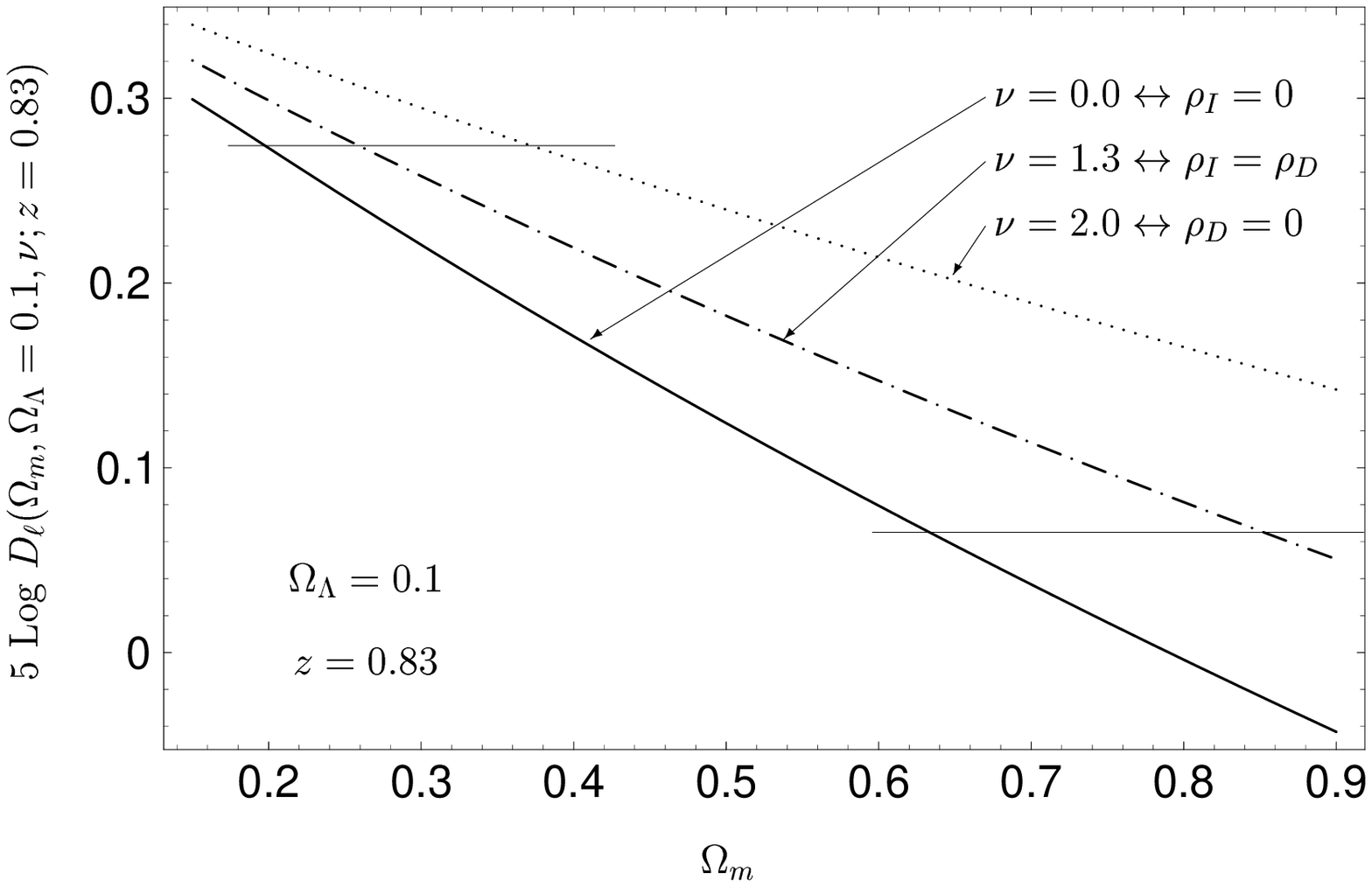,height=80mm,width=120mm}
\vskip -8mm
\caption{Magnitude, $5\log_{10}{H_0\over c}D_{\ell}(\OM,\OL=0.1,\nu;z=0.83)$,
\,as a function of $\OM$ for 
three values of $\nu$.}
\end{minipage}
}
\vskip 8mm
\centerline{
\begin{minipage}[t]{120mm}
\epsfig{figure=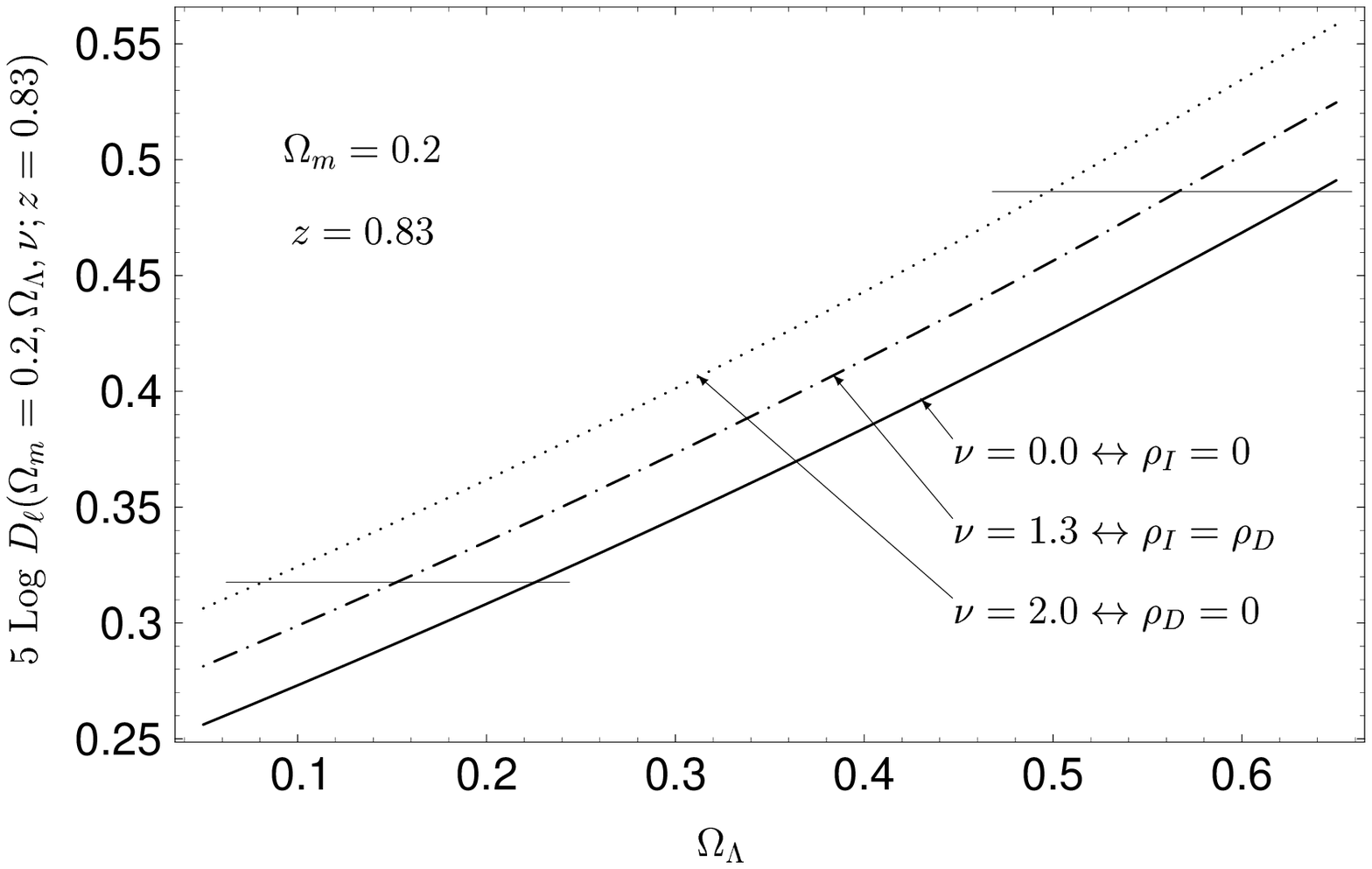,height=80mm,width=120mm}
\vskip -8mm
\caption{Magnitude, $5\log_{10}{H_0\over c}D_{\ell}(\OM=0.2,\OL,\nu;z=0.83)$,
\,as a function of $\OL$ for 
three values of $\nu$.}
\end{minipage}
}
\end{figure}
\section{Concluding Remarks}
Numerous arguments have been made since the '60s against the existence of any effect on  
apparent magnitudes such as given here (see \cite{KR2} for current refs. to dissenting opinions). 
These are almost all weak-lensing arguments, valid as long as density perturbations 
don't produce multiple images (or absorb photons), \ie the FLRW result will coincide with the theoretical
mean of weak lensing observations.  Even if weak lensing arguments can be extended to 
strong lensing perturbations, \eg by not resolving separate images, the theoretical mean of a 
distribution of magnitudes may be of little use in determining $\OM$ or $\OL$. If mass is 
as inhomogeneous as luminosity, the distribution
of magnitudes is expected to be so skewed as to make  the mean statistically insignificant. 
The `most likely' value should be far more useful in a determination of the cosmic parameters.
However, to determine the most likely \mz not only requires knowledge of the average mass density $\rho_0$, 
it requires the modeling of galaxies masses etc. 
What can be determined
by fixing a single additional parameter $\nu$ (which proportions total FLRW mass density 
into  an intergalactic  component and a galactic component)  is the `intergalactic' Hubble curve. 
This \mz can be much more useful in determining $\OM$ or $\OL$ than the 
mean Hubble curve, \eg if galaxies are not larger than they appear optically, the distribution of magnitudes 
(at a given redshift) is expected to peak much closer to the intergalactic value than to the mean value (see
the numerical work \cite{HD}). 
Additionally the effects of galaxy lensing can be controlled by simple selection. All that is required is 
that the observed SNe Ia are separated into 
those with foreground galaxies and those without. Those without should be fit to the intergalactic 
Hubble curve and 
those with (when any are found) could be included by correcting for lensing and/or by averaging in 
with the others to see if the weak-lensing FLRW value can be obtained. Only with enough unbiased and absorption 
corrected data will
the FLRW Hubble curve be useful.

The intergalactic Hubble curve contains  the additional parameter $\nu$; however, if luminous matter
is essentially the whole story most matter is in galaxies and one can put $\nu=2$ as a good approximation. 
Use of the intergalactic Hubble curve is then, in principle, no more involved than use of the standard 
FLRW Hubble curve. The division of $\rho_0$ into  galactic and intergalactic parts for 
gravitational-optics purposes seems simplistic but it is certainly less simplistic than ignoring 
optical effects of inhomogeneities altogether as is done by using FLRW.

\noindent{\bf Further info e-mail:} kantowski@mail.nhn.ou.edu\  {\bf web:} http://www.nhn.ou.edu/$\sim$ski/
\end{document}